\documentclass[preprint,review,12pt]{elsarticle}

\usepackage{amssymb}
\usepackage{amsmath}

\usepackage{hyperref}
\hypersetup{
    colorlinks=true,
    linkcolor=black,
    filecolor=black,      
    urlcolor=blue,
    citecolor=black,
    pdftitle={De-Individualizing fMRI},
    pdfpagemode=FullScreen,
    }
\usepackage[nameinlink]{cleveref}
\usepackage{longtable}
\usepackage{booktabs}
\usepackage{algorithm}
\usepackage{algorithmicx}
\usepackage{algpseudocode}
\algrenewcommand\algorithmicrequire{\textbf{Input:}}
\algrenewcommand\algorithmicensure{\textbf{Output:}}
\usepackage{subcaption}
\usepackage{wrapfig}
\usepackage{braket}

\DeclareMathOperator{\tr}{tr}
\newcommand{\R}{{\mathbb{R}}}

\begin{document}

\begin{frontmatter}

\title{De-Individualizing fMRI Signals via Mahalanobis Whitening and Bures Geometry}

\author[UNC-math]{Aaron Jacobson}
\author[UNC-psychiatry]{Tingting Dan}
\author[UNC-psychiatry,Basel]{Martin Styner}
\author[UNC-psychiatry]{Guorong Wu}
\author[UNC-math]{Shahar Kovalsky}
\author[UNC-math]{Caroline Moosm\"uller}

\affiliation[UNC-math]{organization={Department of Mathematics, University of North Carolina at Chapel Hill},
            addressline={120 E. Cameron Avenue}, 
            city={Chapel Hill},
            postcode={27599}, 
            state={North Carolina},
            country={United States of America}
            }
\affiliation[UNC-psychiatry]{organization={Department of Psychiatry, University of North Carolina at Chapel Hill},
            addressline={101 Manning Dr}, 
            city={Chapel Hill},
            postcode={27514}, 
            state={North Carolina},
            country={United States of America}
            }
\affiliation[Basel]{organization={Department of Biomedical Engineering, University of Basel},
            addressline={Hegenheimermattweg 167}, 
            city={Allschwil},
            postcode={CH-4123}, 
            country={Switzerland}
            }

\begin{abstract}
Functional connectivity has been widely investigated to understand brain disease in clinical studies and imaging-based neuroscience, and analyzing changes in functional connectivity has proven to be valuable for understanding and computationally evaluating the effects on brain function caused by diseases or experimental stimuli. By using Mahalanobis data whitening prior to the use of dimensionality reduction algorithms, we are able to distill meaningful information from fMRI signals about subjects and the experimental stimuli used to prompt them. Furthermore, we offer an interpretation of Mahalanobis whitening as a two-stage de-individualization of data which is motivated by similarity as captured by the Bures distance, which is connected to quantum mechanics. These methods have potential to aid discoveries about the mechanisms that link brain function with cognition and behavior and may improve the accuracy and consistency of Alzheimer's diagnosis, especially in the preclinical stage of disease progression.
\end{abstract}

\begin{keyword}
fMRI working memory tasks \sep Dimensionality reduction \sep Functional connectivity \sep Bures geometry

\end{keyword}

\end{frontmatter}


\begin{figure}[H]
    \centering
    \includegraphics[width=.9\linewidth]{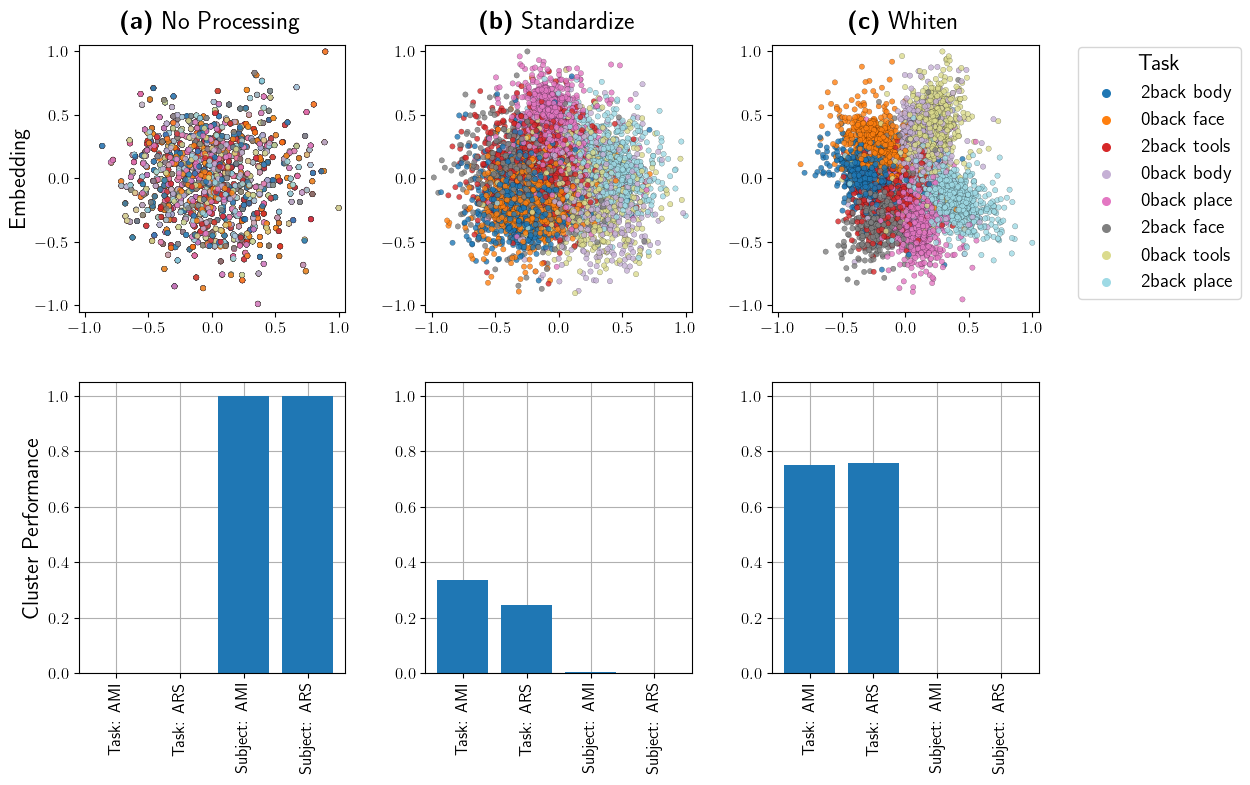}
    \caption{A comparison of embedding quality and clustering performance for various levels of preprocessing. The top row contains embeddings of task scans, and the bottom row displays clustering-based evaluation scores. \textbf{(a)} An embedding based on un-processed task scans. \textbf{(b)} An embedding based on task scans which have been de-meaned and scaled to have a variance of 1. \textbf{(c)} An embedding based on whitened task scans, which gives the best task-based clustering result.}
    \label{fig:SummaryPlot}
\end{figure}

\begin{table}[H]
    \centering
    \begin{tabular}{c|c|c|c}
         & No Preprocessing & Standardize & Whiten \\ \midrule
        AMI: Task & -0.00155 & 0.33409 & 0.75159 \\
        ARS: Task & -0.00089 & 0.24539 & 0.75720 \\
        AMI: Subject & 1.0 & 0.00195 & -0.00248 \\
        ARS: Subject & 1.0 & 0.00082 & -0.00068 \\
    \end{tabular}
    \caption{Quantifications of clustering performance of task scans by individual and by task label. These values correspond to the bars plots in the bottom row of \Cref{fig:SummaryPlot}.}
    \label{tab:SummaryPlotTable}
\end{table}

\section{Introduction}    
    Functional magnetic resonance imaging (fMRI) provides a non-invasive way to examine human brain activity. This imaging technique is often referred to as blood oxygenation level dependent (BOLD) imaging \cite{ogawa1990brain} because it measures changes of cerebral blood oxygenation that are closely related to neuronal activity \cite{logothetis2001neurophysiological}. 
    The analysis of fMRI data (both resting-state and task-based) in neuroscience has become increasingly popular, and it shows promise for detection of neurological disorders, quantification of general intelligence, and identification of brain activity such as is induced by the performance of various memory tasks. In this paper, we take particular interest in the latter of these, as accurate identification of memory-based neural activity may enable accurate identification of issues with memory function, as are present in diseases such as Alzheimer's disease. Future extensions of this work will aim to further tailor the proposed method for the application of neurodegenerative disease diagnosis, seeking to identify new, reliable biomarkers of disease development.
    
    We take as our primary goal the unsupervised clustering of BOLD signals, as measured by task-fMRI, (1) by individual and (2) by the memory task performed by subjects during fMRI scans. We offer a flexible, manifold-learning-based model which accepts BOLD signals as input and admits a distance function as a parameter. The choice of distance function is crucially important, as it significantly affects the meaning of our model's output. We demonstrate that informed choices of distance function enable high-accuracy prediction of either of our target labels (individual or task). For each target label, we offer a meaningful choice of distance and explain its suitability for this problem.
    
    \subsection{Related Works}
        \subsubsection{Similar Data Processing}
            For the identification of task labels, we use a process known as Mahalanobis whitening, which is defined in \Cref{subsec:task_clustering}. The application of Mahalanobis whitening to BOLD data, or multivariate timeseries data in general, is not new. We offer a comparison with two methods which make use of a similar whitening procedure in their analysis of similar data.

            The first is a method for predicting general intelligence from fMRI data by use of recurrent neural networks. Vieira et al. \cite{Vieira2021} uses ZCA whitening to propagate saliency to decorrelated inputs of their proposed model. That is, it uses ZCA in the process of evaluating which model inputs are most important and nowhere else. In contrast, our method makes no use of recurrent neural networks, or in fact, any deep learning architecture. Furthermore, we apply whitening as a pre-processing step to our method rather than in the analysis thereof.
        
            The second is a model which seeks to enhance feature extraction for the purpose of ADHD diagnosis. Aradhya et al. \cite{Aradhya2018} uses a model which regularizes data and subsequently applies Mahalanobis whitening to preprocess the data before applying the spatial filtering method (SFM). They title this process R-SFM (regularized SFM), which is used to process fMRI training data as part of their proposed method for feature extraction. Aradhya et al. then use PBL-McRBFN, a Radial Basis Function (RBF) neural network, to predict classifications, using their extracted features as inputs. Here, whitening is used as a preprocessing step, but it is preceeded by a regularization step and followed by the use of SFM to orthogonally project data. We opt for a simpler pipeline.
            
            Whereas these works use neural networks, we use graph-based manifold learning methods. Furthermore, only one of these works uses whitening as a data preprocessing step, and there, it is embedded in a more complex pipeline than we use. Finally, the aims of these papers are unique from ours, and neither offers an explanation of whitening as de-individualization.
            
        \subsubsection{Manifold Learning Models}
            Of the models which use manifold learning techniques to analyze fMRI data, not all seek to identify or interact with manifolds of BOLD signals. For example, \cite{Wu_Manifold} uses Laplacian-based manifold learning methods to understand manifolds of brain regions, where points on the evaluated manifolds correspond to regions of the brain or networks thereof. In contrast with methods such as these, we seek to identify manifolds on which each point represents a BOLD signal, which provides an evaluation of the relationship between data instances rather than between regions of the brain within a single data instance.

            Other works do consider manifolds of data instances, but generally, they perform different tasks than our model. For example, \cite{Casanova_Manifold} uses manifold learning methods to cluster data from two sources: the Human Connectome Project (HCP), which we use, and the Wake Forest School of Medicine Aging and Alcohol Consumption Database, which we do not use. \cite{Casanova_Manifold} aims to distinguish resting-state data from task-based data, whereas we identify tasks or individuals in a fully task-based setting and use different distances for the comparison of data. Furthermore, we provide a rationale for each choice of distance, seeking explicitly to provide interpretable methods and results.

        \subsubsection{Same Dataset, Different Method}
            A number of works examine the same (or very similar) types of data but use approaches which are different from ours. For example, \cite{Zhang2016_RestVsTask} uses a dictionary learning method to distinguish between task and resting state fMRI from an earlier HCP data release. Another example is \cite{Huang2017_HCP_DeepLearning}, wherein deep convolutional autoencoders (DCAE) are used to evaluate motor-function-related data from the HCP. Two primary features distinguish our work from these. First, while each of these papers examine imaging data from the same source, they focus on different classification tasks and use a different subset of the data available from the HCP. The second difference is in the methods presented; though these papers also perform dimensionality reduction on neuroimaging data, they rely on deep learning methods. In contrast, we use graph-based manifold learning methods for dimensionality reduction; these methods operate in an unsupervised setting, and in general, their design is unique from that of deep neural networks.

\section{Methods}
    \subsection{Dimensionality Reduction}
        Data which exists in a high-dimensional data space often lies on or near a (potentially nonlinear) subspace which is of lower dimension. This assumption, evaluated in \cite{Fefferman2016}, is known as the manifold hypothesis, and it is crucial to the field of dimensionality reduction. When data is distributed on a linear structure, this justifies the use of principal component analysis to project data from its original space to a smaller subspace. When data is distributed along a nonlinear structure, however, more robust methods are required to provide meaningful dimensionality reduction of data. Methods in this class of more powerful methods are known as manifold learning algorithms.
        
        A manifold is a topological space which has the property that, near every point, the space is locally Euclidean. Intuitively, for a given data space, this definition refers to a subset of that data space which has a well-defined tangent space at each point. This is analogous to the sense in which curves can have tangent lines at each point, but this statement generalizes to higher dimensions.

        Manifold learning algorithms seek to uncover and ``flatten'' the manifolds which are assumed to underlie data without causing them to fold or overlap with themselves. They often use graph-based approaches to capture nonlinearity in data structure before using eigendecompositions of distance or Laplacian matrices to embed data in low-dimensional space. In this paper, we use the Isomap algorithm introduced by \cite{Tenenbaum_Isomap} to reduce data dimension. Popular alternatives include t-SNE, UMAP, and Laplacian Eigenmaps \cite{vanDerMaaten2008_tSNE, McInnes_UMAP, Belkin_Eigenmaps}; these methods function differently but have the same goal of embedding nonlinear data structures in low-dimensional spaces. Unless otherwise stated, we will use Isomap for the embeddings we present. For reference, the Isomap algorithm is as follows:

        \begin{algorithm}[H]
                \caption{Isomap \cite{Tenenbaum_Isomap}}
                \begin{algorithmic}[1]
                    \Require Data objects $x_n$, distance $d(\cdot, \cdot)$, number of neighbors $k$
                    \vspace{.25cm}
                    \State Compute pairwise distance matrix $D$ with $D_{i,j} = d(x_i,x_j)$
                    \State Compute the $k$ nearest neighbor graph of the data objects using $D$
                    \State Compute a new pairwise distance matrix $\hat{D}$ where 
                    \[ \hat{D}_{i,j} = \begin{cases}
                        D_{i,j} & \text{nodes $i$ and $j$ are directly connected} \\
                        FW(x_i, x_j) & \text{otherwise}
                    \end{cases}\]
                    where $FW(x_i, x_j)$ is the distance from $x_i$ to $x_j$ on the graph, per the Floyd-Warshall algorithm
                    \State Compute centered Gram matrix $B = -\frac{1}{2}J \hat{D}^{2} J$
                    where the exponent is applied element-wise and $J = I - \frac{1}{n} \vec{1} \vec{1}^{T}$ (a centering matrix)
                    \State Compute eigendecomposition $B = V \Lambda V^{-1}$
                    \vspace{.25cm}
                    \Ensure Embedding coordinates $X = V_k \Lambda_k^{1/2}$, a projection onto a truncated eigenspace of $B$
                \end{algorithmic}
            \end{algorithm}

            We note that the choice of distance function $d$ is crucial to this algorithm's performance. We expect a manifold embedding which is based on a particular distance to place data objects near one another if they are similar in the sense captured by that distance. Different distances capture radically different notions of similarity or dissimilarity, and as such, the output of the Isomap algorithm varies significantly with this choice. By contrast, the parameter $k$ determines the scale which is considered ``local,'' which has an effect on the algorithm's output, but does not affect the meaning of the output in the same way as choice of distance. In \Cref{sec:Results}, we provide two useful choices of distance and a explanation of their meaning.
        
    \subsection{Unsupervised Clustering Evaluation}
        \label{subsec:clustering_evaluation}
        Given that we are interested in clustering data, it is important to consider the ways in which the quality of a clustering algorithm's output can be evaluated. Regardless of whether ground truth labels are available, metrics such as silhouette score (\cite{Rousseeuw1987_SilhouetteScore}) can be used to evaluate how well clusters are separated from one another. 
        
        When ground truth labels are available for data, more metrics are available, including the adjusted Rand score (ARS) and adjusted mutual information (AMI) introduced by \cite{Hubert1985_AdjRandScore} and \cite{JMLR:v11:vinh10a_AdjMutualInfo}, respectively. Each of these metrics accepts two vectors of labels, one containing ground truth labels and the other containing labels assigned by a clustering algorithm, and they output scores which reflect the degree of similarity between the labelings. 
        
        Importantly, ARS and AMI evaluate clustering performance on the basis of data partitioning rather than label alignment. In supervised settings where labels are expected to align with ground truth labels, metrics such as accuracy, precision, recall, or specificity may be reasonable to use. However, our method is unsupervised, and as such, the labels it produces serve to separate data into classes but not align these classes with those provided by ground truth labels. This does not significantly affect the utility of our method, but it does restrict our evaluation of clustering accuracy to those methods which are based on data partitioning and not label alignment, such as ARS and AMI.
        
        ARS and AMI are related to the un-adjusted Rand score and mutual information, which do not have expected values of 0 and do not have comparable meaning between settings with different numbers of classified objects. In contrast, each of these adjusted scores has been modified so that, when evaluated on random labelings, the expected value of their score is 0. Furthermore, the adjusted scores have been normalized so that they have a maximum value of 1, achieved when the algorithmically assigned labels partition data in exactly the same way as the ground truth labels. For the interpretation of our results, it is important to emphasize these facts. To demonstrate why this is the case, we consider a 8-class clustering problem. Given a set of randomly assigned labels, we \textit{do not} expect ARS or AMI to return a value of $\frac{1}{8}$; instead, we expect a score of $0$ from each metric. This means that a statement such as ``AMI$=.125$'' is not equivalent to ``this clustering is $12.5\%$ accurate'' because an AMI of $.125$ indicates better-than-random clustering, whereas a $12.5\%$ accuracy does not.
        
    \subsection{Dataset}
        We use data from the Human Connectome Project (HCP) - Working Memory which contains 960 fMRI time series from scans of unique subjects, each of whom performed 8 memory tasks. 
        Each fMRI scan consists of 393 time points. For each subject, two scans were available: left-to-right (LR) and right-to-left (RL). One scan was used for the test analysis, and the other for the retest analysis. The working memory task involved 2-back and 0-back task conditions for body, place, face, and tool stimuli, along with fixation periods. A resting-state period followed two sequential cognitive task periods in an alternating fashion. For test data, the task events are 2bk-body, 0bk-face, 2bk-tool, 0bk-body, 0bk-place, 2bk-face, 0bk-tool, 2bk-place, while 2bk-tool, 0bk-body, 2bk-face, 0bk-tool, 2bk-body, 2bk-place, 0bk-face, 0bk-place for retest data, respectively. We use minimally processed HCP data, which has undergone distortion correction and was warped to standard space \cite{glasser2013minimal}. To remove motion signal artifacts related to head motion, the ICA-AROMA method \cite{pruim2015ica,pruim2015evaluation} was applied, which considered temporal and spatial features in the fMRI data. Subsequent preprocessing steps involved band-pass filtering the data (0.009–0.08 Hz) and performing regression with mean tissue signals (GM, WM, and CSF), the six movement parameters, and their derivatives. The brain was parcellated into 268 brain regions using the Shen functional atlas \cite{shen2013}, and the residual fMRI signals from all voxels in each parcel were averaged. Where relevant, functional brain networks (functional connectomes) were created by calculating cross-correlations between each pair of network nodes (brain regions). 
        
\begin{figure*}[h]
            \centering
            \begin{subfigure}[t]{0.48\textwidth}
                \vskip 0pt
                \centering
                \includegraphics[height=1.8in, clip, trim=0 0 4.25cm 0]{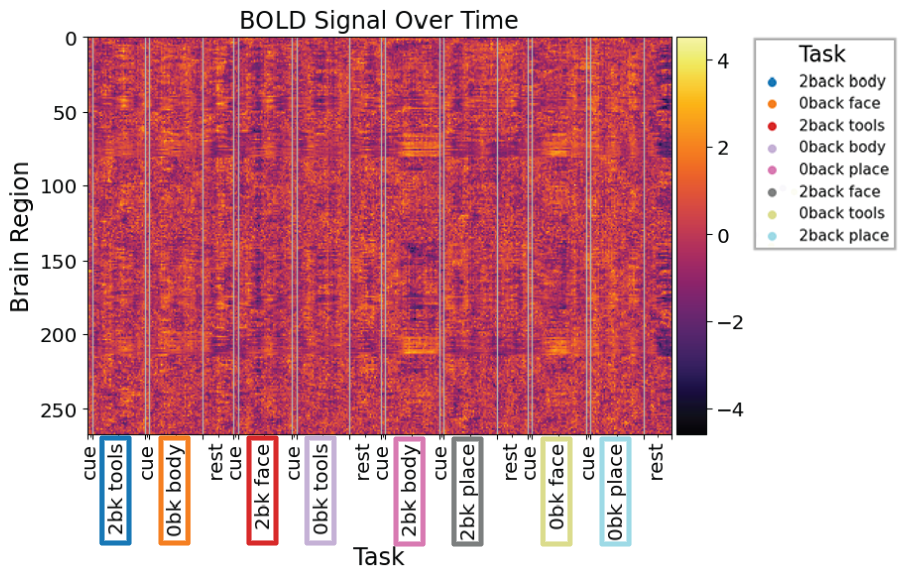}
                \caption{}
                \label{fig:ParentAndTaskScans1}
            \end{subfigure}%
            ~\hspace{.02\linewidth}
            \begin{subfigure}[t]{0.48\textwidth}
                \vskip .075in
                \centering
                \includegraphics[height=1.35in]{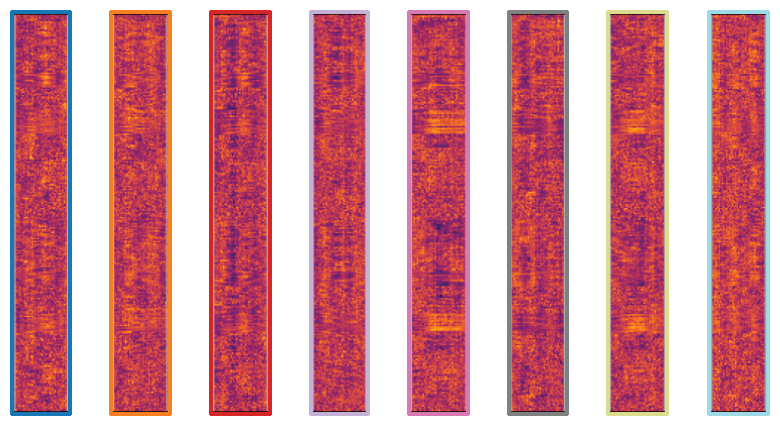}
                \vskip 36pt
                \caption{}
                \label{fig:ParentAndTaskScans2}
            \end{subfigure}
            \caption{One subject's BOLD timeseries after mapping from voxel representation to region-averaged values. \textbf{(a)} A parent scan. Vertical lines delineate tasks, cues, and rests. Extracted tasks are indicated by colored boxes around their labels. \textbf{(b)} 8 task scans extracted from the parent scan. Border color corresponds to the colored boxes around task names in the parent scan.}
        \end{figure*}

        The data has been mapped from its original voxel-based time series representation to one which gives a averaged BOLD value for each region in the Shen-Constable atlas (268 regions). One subject, for whom BOLD-based correlation matrices were defective, was omitted, leaving 8 sub-series for each of 959 subjects in the data set used. We work primarily with these sub-series, which are tasks extracted from an individual's full BOLD time series, and we refer to these as ``task scans,'' whereas we refer to the time series from which they are extracted as ``parent scans.'' See \Cref{fig:ParentAndTaskScans1} and \Cref{fig:ParentAndTaskScans2} for an example of a parent scan and the task scans which are extracted from it.

\section{Results}
    \label{sec:Results}
    The HCP dataset we use includes labels indicating which scans correspond to individual subjects (one label per parent scan) and labels which indicate the task being performed during each sample in the scan (for each parent scan, one label per point in time). Because task scans are sets of contiguous samples with identical task labels, each task scan can be associated with a single individual label and a single task label. In \Cref{subsec:individual_clustering}, we focus on recovering the first of these labels, and in \Cref{subsec:task_clustering}, we recover the second.
    
    \subsection{Clustering by Individual}
        \label{subsec:individual_clustering}
        Growing evidence in neuroscience suggests that the human brain functions as a highly complex distributed system, with spontaneous brain activity occurring near a phase transition that optimizes information transfer, storage, and processing. In this context, it is reasonable to analyze brain activity throughout the brain as a stochastic system. For this reason, we contextualize BOLD signals as random variables and borrow tools from quantum mechanics to compare fMRI scans, drawing an analogy between brain region activation and quantum systems.
        
        \begin{wrapfigure}{r}{0.4\linewidth}
            \centering
            \includegraphics[width=\linewidth]{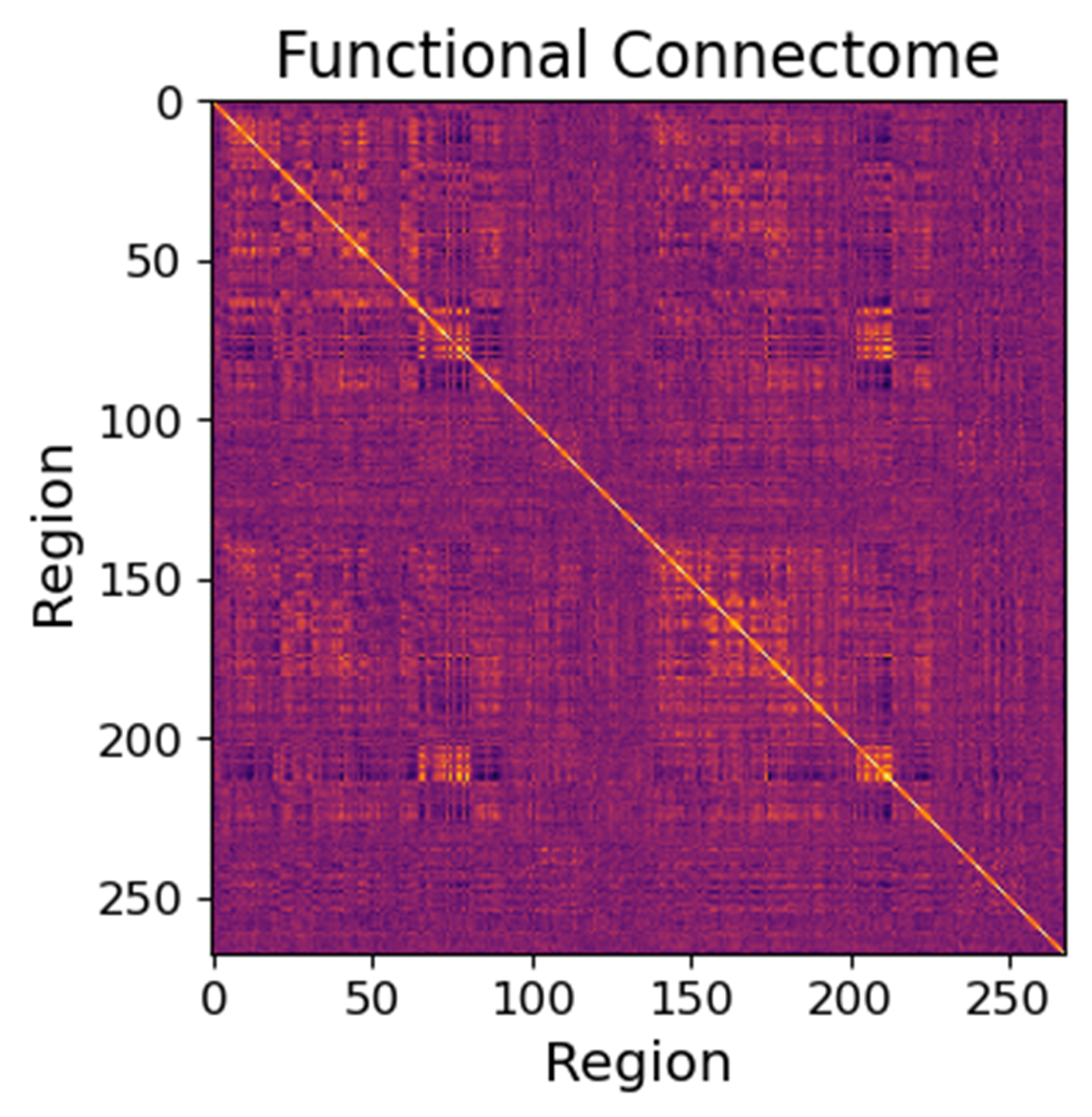}
            \caption{An example of a functional connectome, which is a region-against-region correlation matrix of BOLD signals.}
            \label{fig:FC_example}
        \end{wrapfigure}

        If we interpret the BOLD value of each region as a random variable, we can summarize the relationships between regions as a Pearson's correlation matrix, where correlation is taken region-against-region. This is a standard practice in computational neuroscience, and such a correlation matrix (or in some cases, a covariance matrix) is referred to as a ``functional connectome,'' which describes the pairwise similarity or dissimilarity in the behavior of brain regions. See \Cref{fig:FC_example} for an example. A strong analogy can be made between the functional connectome and an object known as a density matrix, which describes the coherence or interference of basis states in a quantum system. In short, functional connectomes and density matrices are both positive semidefinite (PSD) matrices, the diagonal of each matrix expresses the relative contribution or importance of each basis element (region activity or quantum basis state), and the off-diagonal elements of each express the similarity or dissimilarity of those basis elements.

        The natural distance for comparing density matrices is the Bures distance \cite{Bures1969, Bhatia}, which is built on a notion of similarity known as ``fidelity,'' and when applied to density matrices, this distance measures how distinguishable two quantum systems are. Similarly, when applied to functional connectomes, this distance measures how distinguishable two fMRI scans of brain activity are by the pattern of similarity between brain regions. This distance is applicable to PSD matrices of size $n \times n$ - in fact, it is the geodesic distance on the manifold of PSD matrices - and it is defined as
        \[ d_B(A,B) = \sqrt{\tr(A) + \tr(B) - 2 \sqrt{F(A,B)}} \]
        where $F(A,B)$ is fidelity and is defined by
        \[ F(A,B) = \left[ \tr\left( \left({A^{1/2} B A^{1/2}} \right)^{1/2} \right) \right]^2 \]
        and exponents of $\frac{1}{2}$ applied to matrices refer to matrix square roots (rather than element-wise square roots). For more details about the Bures distance and its use in quantum mechanics, see \Cref{appendix:Bures}.
        
\begin{figure}[H]
            \centering
            \includegraphics[width=0.9\linewidth]{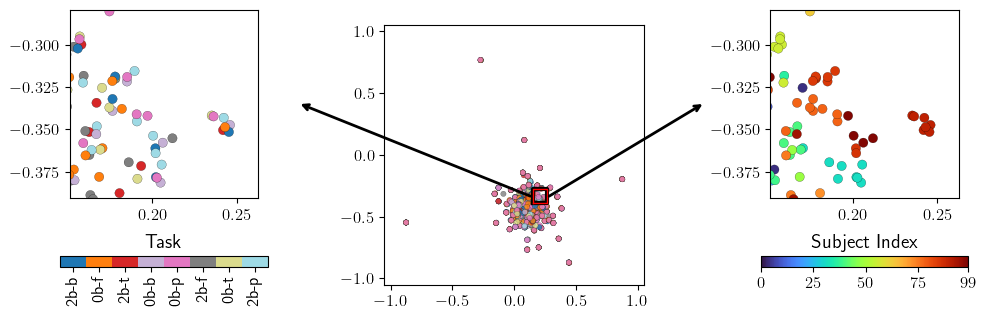}
            \caption{A UMAP embedding of task scans based on pairwise Bures distances between functional connectomes. The smaller figures show the same, zoomed-in view of the data. On the left, task scans are colored by task, and on the right, they are colored by individual. Note that a subset of 100 subjects are shown here, rather than the full dataset.}
            \label{fig:Bures_UMAP_embedding}
        \end{figure}

        We now present an embedding of the task scans from the dataset using pairwise Bures distances between the functional connectomes constructed from task scans; see \Cref{fig:Bures_UMAP_embedding}. Here, we use the UMAP embedding algorithm rather than the Isomap algorithm, as it provides stronger clustering accuracy in this case. Here, each task scan is converted to a functional connectome matrix, pairwise Bures distances are evaluated, and Isomap assigns each task scan to a location in the embedding space. Thus, each point in the embedding represents a single task scan. In this embedding, task scans form clusters based on the individual the task scan came from. This indicates that, in the sense measured by the Bures distance, the stochastic systems represented by the functional connectomes are more similar for task scans from the same individual than for those from different individuals. 

        \begin{table}[H]
            \centering
            \begin{tabular}{c|c|c}
                 & Task & Subject \\ \midrule
                AMI & -0.02747 & 0.7695 \\
                ARS & -0.0007 & 0.6837 \\
            \end{tabular}
            \caption{Clustering performance evaluations for the embedding in \Cref{fig:Bures_UMAP_embedding}.}
            \label{tab:Bures_UMAP_embedding}
        \end{table}

        To evaluate the quality of this embedding method and quantify the degree to which it clusters task scans by individual-based labels, we apply $k$-means clustering to the embedded data, with $k$ set to the number of individuals (i.e. the number of classes in the ground truth labels). Furthermore, we apply this clustering to a 20-dimensional embedding, rather than to the 2-dimensional embedding shown figures such as \Cref{fig:Bures_UMAP_embedding}, so that more structure may be retained for the clustering step. This still represents a significant reduction in dimensionality, as the task scans start as elements of a space which has thousands of dimensions. After applying $k$-means clustering to the data, we compare the clustering labels to ground truth labels using ARS and AMI. For the sake of comparison, we also apply $k$-means clustering with $k$ set to 8 (the number of unique tasks) and provide a clustering analysis for this labeling as well (comparing to ground truth task labels rather than individual labels). We observe that the clustering does not partition task scans by task, but it does strongly partition data by subject. Recall from  \Cref{subsec:clustering_evaluation} that an AMI score of 0.7695 does not directly translate to the statement that the clustering is 76.95\% accurate; instead, this indicates an even stronger level of performance.

        While the Bures distance clusters task scans by individual on the basis of similarity of stochastic behavior, we are also able to cluster task scans by individual using the Frobenius distance between them. This is possible because of small, but consistent, differences in statistical values of BOLD signals from unique individuals. Statistical differences such as a difference in mean BOLD value are easily detected by the Frobenius distance, which is the natural extension of the Euclidean distance to matrices of arbitrary size. For reference, the Frobenius distance between two matrices $A$, $B$ in $\R^{m \times n}$ is given by
        \[ d_{F}(A, B) = \|A-B\|_{F} = \sqrt{\sum_{i,j} (A-B)_{i,j}^{2}} \]
        To demonstrate why this distance is sensitive to differences in mean value, we consider two task scans $A$ and $B$ for which every $A_{i,j} = B_{i,j} + c$ for some fixed $c$. Even though the information contained in the two task scans is in a sense the same, the Frobenius distance between them is nonzero and can potentially be quite large.

\begin{figure}[H]
            \centering
            \includegraphics[width=0.9\linewidth]{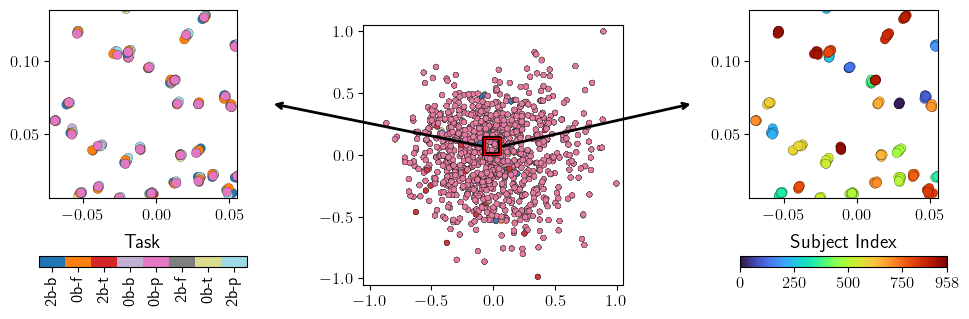}
            \caption{A Isomap embedding of task scans based on pairwise Frobenius distances between un-processed task scans. Colors in the rightmost subplot appear duplicated due to the number of classes present; each cluster of a single color contains all 8 task scans from a single individual.}
            \label{fig:Frobenius_Isomap_Embedding}
        \end{figure}

        We now present an embedding and clustering analysis of Isomap embeddings produced by pairwise Frobenius distances between task scans; see \Cref{fig:Frobenius_Isomap_Embedding} for a 2-dimensional visualization. We observe that individual-based clustering is even stronger than is achieved by the Bures distance, attaining an AMI and ARS score of 1.00, indicating a perfect partitioning of the data by individual. We remark that, while this is a very strong result, there are situations in which the embedding based on Bures distance may still be preferable. For example, large enough deviations between machines used to produce fMRI scans could induce a difference in the mean of measured BOLD signals, which has more to do with the sensor than the actual brain activity being measured. While the Frobenius distance between task scans may be affected by such differences, the Bures distance between functional connectomes will not. This is because the Bures distance is evaluated on correlation matrices, and the construction of correlation matrices eliminates differences in mean and standard deviation which were contained in the original task scans. 
        
        In fact, we do observe statistical differences in BOLD values from task scans which are from different individuals; \Cref{tab:BOLD_Statistical_Vals} contains an example. This may reflect a difference between individuals or a difference in the sensors used to measure BOLD signals. As such, we present this as a strong clustering result, albeit with a dependence on measurement consistency. Furthermore, this result significantly informs the approach taken in \Cref{subsec:task_clustering}.

        \begin{table}[H]
            \centering
            \begin{tabular}{c|c|c}
                 & Task & Subject \\ \midrule
                AMI & -0.00155 & 1.0 \\
                ARS & -0.00089 & 1.0 \\
            \end{tabular}
            \caption{Clustering performance evaluations for the embedding in \Cref{fig:Frobenius_Isomap_Embedding}.}
            \label{tab:Frobenius_Isomap_Embedding}
        \end{table}

        \begin{table}[H]
            \centering
            \begin{tabular}{c|c|c|c|c}
                Subject / Task & Minimum & Mean & Maximum & Standard Dev.\\
                \midrule
                101 / 2bk tools & 3567.50 & 10315.86 & 15208.00 & 2129.53 \\
                101 / 0bk body & 3562.10 & 10316.64 & 15183.00 & 2128.34 \\
                101 / 2bk face & 3555.00 & 10308.82 & 15191.00 & 2127.56 \\
                \midrule
                103 / 2bk tools & 3108.60 & 10051.69 & 14424.00 & 2184.05 \\
                103 / 0bk body & 3125.10 & 10048.66 & 14454.00 & 2183.18 \\
                103 / 2bk face & 3111.10 & 10051.25 & 14441.00 & 2184.41 \\
            \end{tabular}
            \caption{A table demonstrating the small but consistent differences in statistical values between BOLD signals from different subjects.}
            \label{tab:BOLD_Statistical_Vals}
        \end{table}

    \subsection{Clustering by Task}
        \label{subsec:task_clustering}
        We are also interested in extracting task-based labels from the same data. In fact, extracting this set of labels may prove more useful in medical applications than the extraction of individual-based labels; accurately distinguishing between modes of memory function (such as distinguishing between memory tasks being performed) may enable accurate distinction between healthy and unhealthy memory function, as is at issue in neurodegenerative diseases such as Alzheimer's disease. To accurately distinguish modes of memory function, however, we must control for individual-level differences that may obscure the desired task information. Fortunately, the clustering results presented in \Cref{subsec:individual_clustering} provide insight which informs a two-stage normalization procedure that helps isolate the desired information.

        We recall from \Cref{subsec:individual_clustering} that there is some similarity between the task scans from a single subject in the sense of correlation structure between regions' BOLD signals, and there is also similarity between the task scans from a single subject which depends on statistical differences in BOLD values. These differences in correlation structure and statistical values distinguish task scans by individual rather than by task, and these signals obscure information about the task being performed. Thus, we seek to remove the information contained in these similarities with the goal of removing the dominant feature (individual) and leaving behind the desired feature (task).

        To do this, we apply two transformations to our data prior to distance evaluation. The first is standardization, wherein we remove the mean from each region's BOLD signal over the duration of the parent scan. Given a parent scan $S \in \R^{m \times n}$ (which measures $m$ regions at $n$ points in time) and a ones vector $\vec{1} \in \R^{n}$, we define the new, mean $0$ parent scan by 
        \[ \bar{S} = S - \frac{1}{n}S\vec{1} \]
        This subtracts a constant from each row of the parent scan such that the row has mean $0$, removing the most significant source of statistical difference between individuals' scans. At this step, we could also scale the BOLD values of the de-meaned rows such that they have standard deviation $1$ in order to further reduce statistical similarity which groups data by individual. We note that the the next processing step makes such a scaling redundant.

        The second preprocessing step we apply is Mahalanobis whitening \cite{Kessy02102018_whitening, Bell1997-ZCA_Whitening}, which linearly transforms columns of parent scans (which correspond to points in time) such that the whitened data $S_w$ satisfies $\operatorname{cov}(S_w, S_w) = I$. Such a transformation ensures that correlation structure between regions is erased from the data and the standard deviation of each region after the transformation is equal to $1$. The normalization of standard deviations further removes the individual-based similarity observed in simple statistical values, and destruction of correlation structure removes the individual-based similarity which was revealed by the Bures distance. Given a parent scan $S$, we compute the whitened parent scan as 
        \[ S_w = W^{-1/2}\bar{S} \]
        where $\bar{S}$ is the de-meaned parent scan, $W$ is a whitening matrix chosen to satisfy $\operatorname{cov}(S_w, S_w) = I$, and the exponent refers to a fractional matrix exponent. In general, there are arbitrarily many suitable choices of $W$. In our case, it is natural to choose $W = \operatorname{cov}(\bar{S}, \bar{S})$, which is the covariance-based functional connectome, calculated using de-meaned BOLD signals. As an important note, we apply these transformations \textit{prior} to extracting task scans from parent scans, meaning that all task scans from a single fMRI session are processed together.

        These steps serve to eliminate the observed individual-based similarity from the data. To compare the preprocessed data, we once again use the Frobenius distance. Composing these two transformations with the Frobenius distance produces a dissimilarity function $d_M$. Given two task scans $T_1$ and $T_2$, we define $d_M$ as
        \[ d_M(T_1,T_2) = d_F(T_{1,w}, T_{2,w}) \]
        where $T_{i,w}$ is the task scan $T_i$ extracted from the preprocessed parent scan $S_{i,w}$. This definition is simply a composition of the preprocessing steps, task scan extraction, and the Frobenius distance (in that order). 
        
        The name ``distance'' is, technically, a misnomer; the function breaks the triangle inequality. We do not, however, observe any significant negative effects of this fact in our results. Furthermore, we call this function $d_M$ distance to indicate preprocessing which includes Mahalanobis whitening while avoiding a naming conflict with the Mahalanobis distance, which is related and already has definition elsewhere \cite{Mahalanobis}.

\begin{figure}[H]
            \centering
            \includegraphics[width=0.9\linewidth]{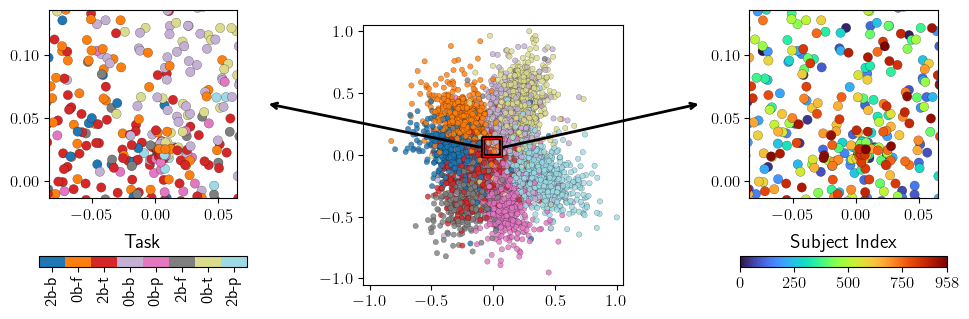}
            \caption{A Isomap embedding of task scans based on pairwise $d_M$ distances between task scans. In contrast with \Cref{fig:Frobenius_Isomap_Embedding}, task scans now cluster by task rather than by subject, leading to significantly improved task-based clustering performance.}
            \label{fig:Preprocessed_embedding}
        \end{figure}

        We now present an Isomap embedding of the dataset which uses $d_M$ as it's pairwise distance function. In \Cref{fig:Preprocessed_embedding}, the embedding of task scans shows visually obvious clustering by task label; \Cref{tab:Preprocessed_embedding} shows corresponding ARS and AMI values for $k$-means clustering. We see that $k$-means clustering does not reveal any meaningful information about individuals' identities, but it does effectively partition data according to task labels. Recall the discussion in \Cref{subsec:clustering_evaluation}; an ARS or AMI score near $.75$ is not directly equivalent to ``$75\%$ accuracy'' but instead expresses an even stronger level of performance. Given that this is an 8-class clustering problem, scores near $.75$ constitute a strong result which demonstrates that this pairwise distance and the corresponding embedding effectively extract the desired information from the dataset.

        \begin{table}[H]
            \centering
            \begin{tabular}{c|c|c}
                 & Task & Subject \\ \midrule
                AMI & 0.75159 & -0.00248 \\
                ARS & 0.75720 & -0.00068 \\
            \end{tabular}
            \caption{Clustering performance evaluations for the embedding in \Cref{fig:Preprocessed_embedding}.}
            \label{tab:Preprocessed_embedding}
        \end{table}

    \subsubsection{Validation Dataset}
\begin{figure}[H]
            \centering
            \includegraphics[width=0.9\linewidth]{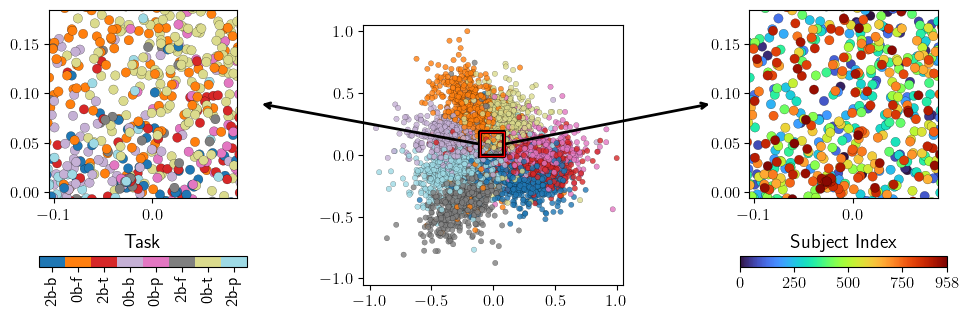}
            \caption{A recreation of \Cref{fig:Preprocessed_embedding} using the validation data set. The same clustering characteristics are evident.}
            \label{fig:Preprocessed_embedding_validation}
        \end{figure}

        The HCP dataset we use also supplies a second parent scan for each individual. In this second parent scan, the same tasks are performed in a new order. This allows us to evaluate whether our model is dependent on the ordering of tasks and further demonstrate the effectiveness of our method. In \Cref{fig:Preprocessed_embedding_validation} and \Cref{tab:Preprocessed_embedding_validation}, we observe even stronger clustering performance than in the original dataset. From this we conclude that our method is robust to changes in the order of tasks in a parent scan.

        \begin{table}[H]
            \centering
            \begin{tabular}{c|c|c|c}
                 & Task & Subject \\ \midrule
                AMI & 0.81359 & -0.00177 \\
                ARS & 0.81974 & -0.00047 \\
            \end{tabular}
            \caption{Clustering performance evaluations for the embedding in \Cref{fig:Preprocessed_embedding_validation}.}
            \label{tab:Preprocessed_embedding_validation}
        \end{table}

\section{Discussion}
    We have identified multiple distance functions which provide meaningful comparisons of task scans from the HCP Working Memory dataset. We have demonstrated that these distances work well with unsupervised learning methods to successfully partition data either by individual or by task. Furthermore, we demonstrate that knowledge of the type of similarity or dissimilarity measured by various distance functions can provide critical insight into the development of preprocessing pipelines. It is in this way that we connect the individual-based and task-based clustering settings; results that cluster data by individual inform preprocessing steps which remove confounding effects and enable task-based clustering.

    Future work will focus on the extension of this methodology and its application to datasets such as ADNI and OASIS \cite{ADNI, OASIS}, each of which provides data with labels related to Alzheimer's diagnosis. We will seek to remove confounding effects in this data to isolate the desired information, which may be used as a biomarker of Alzheimer's development. 

\section{Acknowledgements}
    Data were provided [in part] by the Human Connectome Project, WU-Minn Consortium (Principal Investigators: David Van Essen and Kamil Ugurbil; 1U54MH091657) funded by the 16 NIH Institutes and Centers that support the NIH Blueprint for Neuroscience Research; and by the McDonnell Center for Systems Neuroscience at Washington University.

    This research was partially supported by a seed grant from the School of Data Science and Society at UNC.

    A.J. and C.M. were partially supported by NSF award DMS-2410140.

    S.K. was partially supported by NSF award DMS-2152289, FRG: Collaborative Research: Mathematical and Statistical Analysis of Compressible Data on Compressive Networks.

\appendix
\section{The Bures Distance}
\label{appendix:Bures}
    Our primary reference in this section is \cite{Bengtsson_QuantumBook}, which discusses most of the quantum-mechanics related content here. We also make use of \cite{Bures1969}, written by Donald Bures, and \cite{Bhatia}, which provides a useful variant of the Bures metric applicable to matrices that do not necessarily have a trace equal to 1.

    The Bures distance comes from quantum mechanics, and in this setting, the object to which it is typically applied is a density matrix (rather than PSD matrices in general). Density matrices represent quantum systems in terms of the relationships between their basis states; mathematically, they are PSD matrices with trace 1. Further constraints exist to ensure that these matrices describe physically realistic systems, but because these further restrictions guarantee physical relevance (rather than whether the Bures distance is well-defined), we will disregard them for now. Given a state $\Psi = \sum_n p_n \ket{\Psi_n}$ of a quantum system, which is a statistical mixture of pure states $\ket{\Psi_n}$ with weights $p_n$ (with $p_n \geq 0$ and $\sum p_n = 1$), we calculate the density matrix as
    \[ \rho = \sum_n p_n \ket{\Psi_n}\bra{\Psi_n} \]
    This may be understood as an inner product of coefficient vectors, where $\ket{\Psi_n}$ is a vector which contains coefficients $c_i$ of the component wave functions of $\Psi$, and $\bra{\Psi_n}$ is the conjugate transpose of that vector. For $\Psi$ to be a pure state, only one $p_n$ will be nonzero (and must equal 1), and the sum notation is no longer necessary.
    
    In our paper, we present the Bures distance in a general form, applicable to all positive semidefinite matrices. When applied only to density matrices, which have trace equal to 1, we can rewrite the Bures distance in a form that appears more often in quantum mechanics literature:
    \begin{gather*}    
    d_B(A,B) = \sqrt{2 - 2 \tr\left( \left({A^{1/2} B A^{1/2}} \right)^{1/2} \right)} = \sqrt{2 - 2 \sqrt{F(A,B)}} \\
    F(A,B) = \left[ \tr\left( \left({A^{1/2} B A^{1/2}} \right)^{1/2} \right) \right]^2
    \end{gather*} 
    Here, $A$ and $B$ represent density matrices of quantum states (in quantum mechanics literature, they are often notated as $\rho, \sigma$, etc.). We note that, for $A,B$ with trace 1, $d_B(A,B) \in [0, \sqrt{2}]$. The function $F(\cdot, \cdot)$ is called ``fidelity,'' and it measures the similarity of two quantum systems; as such, it outputs a higher value when two density matrices are similar and a lower value otherwise. Fidelity, when applied to density matrices, is bounded above by 1. The Bures distance turns the similarity expressed by fidelity into a distance; we can rewrite the Bures distance once more to see why this is the case:
    \[ d_B(A,B) = \sqrt{2 - 2 \sqrt{F(A,B)}} = \sqrt{\left(1 - \sqrt{F(A,B)}\right) + \left(1 - \sqrt{F(A,B)}\right)} \]
    Each term $\left(1 - \sqrt{F(A,B)}\right)$ is equal to 1 when fidelity is 0 and equal to 0 when fidelity is 1. This means that $d_B(\cdot,\cdot)$ outputs higher values when comparing dissimilar matrices and lower values when comparing similar matrices. When the matrices compared do not have trace 1, the form of the distance changes (to the expression presented in \Cref{subsec:individual_clustering}) and the bounds on the distance change, but the fact that the Bures distance is derived from fidelity remains the same.

    \subsection{Comparison of Correlation and Density Matrices}

    We begin by noting that correlation matrices, like density matrices are positive semidefinite. Any $n \times n$ correlation matrix can be converted to a density matrix simply by dividing by $n$. This normalizes the trace of the matrix to be 1, which produces a density matrix (which may or may not describe a physically plausible system). Furthermore, we can represent both density and correlation matrices as matrices of inner products. To do this, we start by writing the form of a density matrix. For simplicity, we do so with a ``pure'' state, for which only one $p_i$ is nonzero (and must equal 1). In this case, the sum notation in the definition of $\rho$ can be eliminated.
    \[ \rho = \ket{\Psi}\bra{\Psi} = \begin{bmatrix}
        c_1 & c_2 & \dots
    \end{bmatrix} \begin{bmatrix}
        c_1^* \\ c_2^* \\ \vdots
    \end{bmatrix} = \begin{bmatrix}
        c_1 c_1^* & c_1 c_2^* & \dots \\
        c_1^* c_2 & c_2 c_2^* & \dots \\
        \vdots & \vdots & \ddots
    \end{bmatrix} \]
    For comparison, we now construct a correlation matrix. Let $T$ be a random variable with values in $\R^d$ with probability measure $\mu(x)$ (e.g. a multivariate time series). Let $T_j$ be the $j$th component of $T$ (e.g. the $j$th variable in the multivariate time series). Then we have
    \[ Corr = \begin{bmatrix}
        \langle \frac{T_1 - \mathbb{E}(T_1)}{\sigma_{T_1}}, \frac{T_1 - \mathbb{E}(T_1)}{\sigma_{T_1}} \rangle & \langle \frac{T_1 - \mathbb{E}(T_1)}{\sigma_{T_1}}, \frac{T_2 - \mathbb{E}(T_2)}{\sigma_{T_2}} \rangle & \dots \\
        \langle \frac{T_2 - \mathbb{E}(T_2)}{\sigma_{T_2}}, \frac{T_1 - \mathbb{E}(T_1)}{\sigma_{T_1}} \rangle & \langle \frac{T_2 - \mathbb{E}(T_2)}{\sigma_{T_2}}, \frac{T_2 - \mathbb{E}(T_2)}{\sigma_{T_2}} \rangle & \dots \\
        \vdots & \vdots & \ddots
    \end{bmatrix} \]

    We observe that both correlation matrices and density matrices can be written as matrices of inner products; for each kind of matrix, this means that elements describe some kind of similarity or dissimilarity between objects (in this example, wave functions and normalized random variables). This makes it clear that, while density and correlation matrices have similar mathematical properties, there is a richer comparison to be made between them. We provide a brief interpretation here.

    Considering first the elements on the diagonal of these matrices, we see that these elements provide information about the scale or size of the objects being compared, as they are inner products of objects with themselves. Comparing elements on the diagonal of these matrices gives information about the relative size or importance of each component being compared. For density matrices, this means that the elements on the diagonal give the probability that a quantum system will be observed in each component state. For correlation matrices, the diagonal gives the relative scale of each variable, but because we consider correlation rather than covariance, each diagonal element is equal to 1. This can be interpreted as a re-scaling of each variable to ensure that they are all equivalently ``important.''

    Considering next the elements off of the diagonal, we see that these inner products give information about the similarity or dissimilarity between components. For density matrices, this is the interference or coherence between basis states. For correlation matrices, this is the Pearson's correlation between variables.
    
    Because density and correlation matrices are similar in both mathematical properties and interpretation, we conclude that the Bures distance is the natural choice of distance function not only for quantum mechanics, but also for analysis of correlation matrices.

 \bibliographystyle{elsarticle-num} 
 \bibliography{refs.bib}

\end{document}